\documentclass[twocolumn, tighten]{aastex61}
\received{July 11, 2017}
\revised{February 14, 2018}
\submitjournal{ApJ}
\shorttitle{Superresolution Interferometric Imaging with Sparse Modeling Using Total Squared Variation}
\shortauthors{K.~Kuramochi et al.}
\definecolor{MyDarkBlue}{rgb}{0,0.08,0.5}
\definecolor{MyDarkRed}{rgb}{0.7,0.02,0.02}
\definecolor{MyDarkGreen}{rgb}{0.0,0.7,0.0}

\newcommand{\nrao}{National Radio Astronomy Observatory, 520 Edgemont Rd, Charlottesville, VA 22903, USA}
\newcommand{\utokyo}{Department of Astronomy, Graduate School of Science, The University of Tokyo, 7-3-1 Hongo, Bunkyo-ku, Tokyo 113-0033, Japan}
\newcommand{\naoj}{National Astronomical Observatory of Japan, 2-21-1 Osawa, Mitaka, Tokyo 181-8588, Japan}
\newcommand{\haystack}{Massachusetts Institute of Technology, Haystack Observatory, 99 Millstone Rd, Westford, MA 01886, USA}
\newcommand{\bhi}{Black Hole Initiative, Harvard University, 20 Garden Street, Cambridge, MA 02138, USA}
\newcommand{\ism}{Department of Statistical Science, School of Multidisciplinary Sciences, Graduate University for Advanced Studies,  10-3 Midori-cho, Tachikawa, Tokyo 190-8562, Japan}
\newcommand{\sokendaiism}{Graduate University for Advanced Studies, 10-3 Midori-cho, Tachikawa, Tokyo 190-8562, Japan}
\newcommand{\perimeter}{Perimeter Institute for Theoretical Physics, 31 Caroline Street, North Waterloo, Ontario N2L 2Y5, Canada}
\newcommand{\asiaa}{Institute of Astronomy and Astrophysics, Academia Sinica, P.O. Box 23-141, Taipei 10617, Taiwan}
\newcommand{\sokendainaoj}{Department of Astronomical Science, School of Physical Sciences, Graduate University for Advanced Studies, 2-21-1 Osawa, Mitaka, Tokyo 181-8588, Japan}
\usepackage{amsmath}
\DeclareMathOperator*{\argmin}{argmin}
\begin{document}
\title{Superresolution Interferometric Imaging with Sparse Modeling Using Total Squared Variation --- Application to Imaging the Black Hole Shadow}
%
\correspondingauthor{Kazunori Akiyama}
\email{kakiyama@mit.edu}
\author{Kazuki Kuramochi}
\affil{\utokyo}
\affil{\naoj}
\author{Kazunori Akiyama}
\altaffiliation{NRAO Jansky Fellow}
\affil{\nrao}
\affil{\haystack}
\affil{\bhi}
\affil{\naoj}
\author{Shiro Ikeda}
\affil{\ism}
\affil{\sokendaiism}
\author{Fumie Tazaki}
\affil{\naoj}
\author{Vincent L. Fish}
\affil{\haystack}
\author{Hung-Yi Pu}
\affil{\perimeter}
\affil{\asiaa}
\author{Keiichi Asada}
\affil{\asiaa}
\author{Mareki Honma}
\affil{\naoj}
\affil{\sokendainaoj}
%
\begin{abstract}
We propose a new superresolution imaging technique for interferometry using sparse modeling,  utilizing two regularization terms: the $\ell_1$-norm and a new function named Total Squared Variation (TSV) of the brightness distribution. TSV is an edge-smoothing variant of Total Variation (TV), leading to reducing the sum of squared gradients. First, we demonstrate that our technique may achieve super-resolution of $\sim 30$\% compared to the traditional CLEAN beam size using synthetic observations of two point sources. Second, we present simulated observations of three physically motivated static models of Sgr A* with the Event Horizon Telescope (EHT) to show the performance of proposed techniques in greater detail. We find that $\ell_1$+TSV regularization outperforms $\ell_1$+TV regularization with the popular isotropic TV term and the Cotton-Schwab CLEAN algorithm, demonstrating that TSV is well-matched to the expected physical properties of the astronomical images, which are often nebulous. Remarkably, in both the image and gradient domains, the optimal beam size minimizing root-mean-squared errors is $\lesssim 10$~\% of the traditional CLEAN beam size for $\ell_1$+TSV regularization, and non-convolved reconstructed images have smaller errors than beam-convolved reconstructed images. This indicates that the traditional post-processing technique of Gaussian convolution in interferometric imaging may not be required for the $\ell_1$+TSV regularization. We also propose a feature extraction method to detect circular features from the image of a black hole shadow with the circle Hough transform (CHT) and use it to evaluate the performance of the image reconstruction. With our imaging technique and the CHT, the EHT can constrain the radius of the black hole shadow with an accuracy of $\sim 10-20$~\% in present simulations for Sgr A*, suggesting that the EHT would be able to provide useful independent measurements of the mass of the supermassive black holes in Sgr A* and also another primary target, M87.  
\end{abstract}
\keywords{accretion, accretion disks --- black hole physics --- Galaxy: center --- techniques: high angular resolution ---  techniques: image processing --- techniques: interferometric}

\section{Introduction}\label{sec:1}
Extremely high angular resolution is a fundamental pursuit in modern observational astronomy.  Finer resolution provides a more detailed picture of astronomical objects and has been essential for breakthroughs in understanding the nature of these objects. Interferometry is one of the most effective approaches for obtaining high angular resolution. With a collection of telescopes, the synthesized aperture provides a nominal resolution (often referred to as ``beam size'' in radio astronomy and ``diffraction limit'' in optical astronomy) of $\theta \approx \lambda/D_{\rm max}$, where $D_{\rm max}$ is the maximum length of the baseline between two telescopes, projected in the plane normal to the direction of observation. The technique of interferometry has been expanded from radio to optical wavelengths in the last decades \citep[e.g.][]{thiebaut2013, thompson2017}.

In the history of astronomy, the highest angular resolution has been pursued with very long baseline interferometry (VLBI), which utilizes intercontinental baselines or even baselines to space.  Recent progress in VLBI technology has opened a new window of VLBI observations at short/sub-millimeter wavelengths ($\lambda \lesssim 1.3$~mm, $\nu \gtrsim 230$~GHz). Ground-based VLBI observations at this wavelength have been realized with the Event Horizon Telescope \citep[EHT;][]{doeleman2009}. The EHT has been achieving an angular resolutions of a few tens of microarcseconds \citep[e.g.][]{doeleman2008,doeleman2012,fish2011,fish2016,lu2012,lu2013,akiyama2015,johnson2015b}, which is the finest angular scale accessible only with the EHT or lower-frequency space VLBI \citep[e.g. RadioAstron;][]{kardashev2013}. The EHT resolves compact structure of the magnetized plasma on scales of a few Schwarzschild radii ($R_s = 2GM/c^2$) in the vicinity of the supermassive black holes (SMBH) in the Galactic Center source Sgr A* \citep{doeleman2008,fish2011,fish2016,johnson2015b} and the nucleus of M87 \citep{doeleman2012,akiyama2015}. Imaging capability with the EHT will be provided in the next few years with the critical addition of new sensitive telescopes such as the Atacama Large Submillimeter/millimeter Array \citep[ALMA; e.g.][]{fish2013}.

In parallel with developments in observational instruments, significant progress has been made on techniques for interferometric imaging to obtain higher-fidelity and/or higher-resolution images in the last decades. In particular, superresolution imaging has been identified as an important goal for the EHT. A practical limit for a ground-based, 1.3~mm VLBI array like the EHT is $\sim 25$ $\mu$as ($=1.3$~mm/$10000$~km), which is comparable to the radius of the black hole shadow in Sgr A* and M87. Previous work has shown that the most widely-used CLEAN algorithm \citep{hogbom1974} and its variants \citep[e.g.][]{clark1980,schwab1984} have difficulty obtaining high-fidelity images for both Sgr A* and M87 \citep{honma2014,chael2016,akiyama2017a,akiyama2017b}. Hence, numerous works have been presented in the recent few years on new imaging techniques for the EHT \citep{honma2014,lu2014,lu2016,fish2014,bouman2016,chael2016,johnson2016c,akiyama2017a,akiyama2017b}.

The authors have developed a series of techniques mainly designed for VLBI named ``sparse modeling'' \citep{honma2014,ikeda2016,obuchi2017,akiyama2017a,akiyama2017b}, utilizing the expected sparsity of the ground truth image. The underlying idea comes from the theory of compressed sensing (also known as compressive sensing and sparse sampling), revealing that an ill-posed linear problem like interferometric imaging may be solved accurately if the underlying solution vector is sparse \citep{donoho2006,candes2006}. The pioneering work on sparse reconstruction techniques for radio interferometry is presented in \cite{wiaux2009a} and \cite{wiaux2009b}. Numerous other papers\footnote{The authors maintain a publicly available list of papers in ADS beta with cooperation of other groups working on sparse imaging techniques for interferometry: \url{https://ui.adsabs.harvard.edu/\#user/libraries/wmxthNHHQrGDS2aKt3gXow}.} have been presented, mostly geared toward data from next-generation low-frequency interferometers \citep{wiaux2010,wenger2010,li2011,mcewen2011,carrillo2012,carrillo2014,garsden2015,dabbech2015,onose2016,onose2017}. Both our and other groups' work have shown that these state-of-the-art sparse reconstruction techniques outperform the conventional CLEAN technique and its variants.

In previous work, we have utilized two convex regularization functions that represent the sparsity of the image (see \S\ref{sec:2} for details).  The first regularizer is the $\ell _1$ norm of the image, which favors sparsity of the image itself. The underlying idea, which comes from LASSO \citep[Least Absolute Shrinkage and Selection Operator;][]{tibshirani1996}, is that minimizing the $\ell_1$ norm of the solution leads to a sparse solution in the image domain \citep[see, e.g.][for details]{honma2014,akiyama2017b}. The second regularizer is the Total Variation (TV) of the image, which favors sparsity in the image \emph{gradient} domain. For the TV term, we have heretofore adopted the isotropic TV \citep{rudin1992}, which has been the most widely-used form for astronomical imaging in previous literature \citep[e.g.][]{wiaux2010,mcewen2011,carrillo2012,carrillo2014,uemura2015,chael2016}. The inclusion of TV is essential to solve for images with multi-scale structures \citep[e.g.~see][]{akiyama2017b}. Our imaging technique utilizing $\ell _1$+TV regularization can handle full complex visibilities in full polarization \citep{akiyama2017b} and also closure phases \citep{akiyama2017a}. Note that the majority of prior work takes an alternative approach, changing the basis of the image to a sparser one using wavelet or curvelet transforms, which has been successful as well.

An important advantage of all new state-of-the-art imaging techniques, which solve the observational equation directly with convex regularization, is the capability of high-fidelity super-resolution imaging beyond the conventional resolution of $\sim \lambda/D_{\rm max}$. This capability is given by both analyticity of data and constraints on Fourier modes of images with regularization functions \citep{narayan1986}. Here we note that, in radio astronomy, the term of "angular resolution" has often been used for two meanings in literature, which we strictly distinguish in this paper; the first one is its literal definition, a minimum separation of two (point) sources identifiable with observations (henceforth {\it effective angular resolution}); another popular usage is for the optimal size of the restoring Gaussian beam used to convolve the raw reconstructed image (henceforth {\it optimal beam size}). For the effective resolution, \citet{honma2014} concludes that superresolution of $\sim 0.25 \lambda /D_{\rm max}$ would be achievable with $\ell _1$-regularization using one-dimensional synthetic observations. For the optimal beam size, our recent work finds that $\ell _1$+TV regularization can achieve an optimal beam size of $\sim 0.2-0.3\lambda/D_{\rm max}$ both in total-intensity (i.e.~Stokes $I$) imaging using visibility amplitudes and closure phases \citep{akiyama2017a} and in full-polarization imaging using full-complex visibilities \citep{akiyama2017b}. 
Although our previous work demonstrates that our sparse reconstruction technique is an attractive choice for high-fidelity imaging, our previous results have been limited by two issues.

The first issue is that edge-preserving reconstruction, which is a representative property of isotropic TV regularization, introduces artificially edgy boundaries with too flattened brightness distributions in reconstructed images for simulated observations of black hole accretion disks and jets with the EHT \citep{akiyama2017a,akiyama2017b}. Such edge-preserving regularization is generally favored for natural images or photographs of the objects in the Earth, where most of the target objects are solid or liquid materials with sharp-edged boundaries. However, in astronomy, observed objects are often diffuse, and their images do not have clear boundaries due to a gradual change in both the emissivity and the absorption coefficient.

The second issue is evaluation of the image fidelity, which has been common in all the past EHT-related imaging papers. Our past work has adopted metrics comparing the reconstructed images with the ground truth image on a pixel-by-pixel basis, such as the Image Residual \citep[][]{honma2014}, Mean Square Error \citep[MSE;][]{lu2014,lu2016,fish2014,bouman2016} or its variant Normalized Root Mean Square Error \citep[NRMSE;][]{chael2016,akiyama2017a,akiyama2017b} and Structural Dissimilarity index \citep[DSSIM;][]{lu2014,lu2016,fish2014,bouman2016}. These metrics tend to be overly sensitive to the brightest regions in the image and may not always be an appropriate indicator of the goodness of feature reconstruction \citep[e.g.~see \S4.1 in][]{akiyama2017b}, which is important for interpreting the images physically.

In this work, we propose a new technique for interferometric imaging using sparse modeling, utilizing both the $\ell_1$-norm and a TV term. For the first issue, we newly adopt an alternative form of TV, named ``Total Squared Variation,'' which prefers smoothed images with less sharp edges. For the second issue, we involve a new metric using the image gradient, which is more sensitive to the size of the emission region and continuity of the emission structure than conventional metrics. In addition, we present a new feature extraction method designed for the black hole shadow using the circle Hough transform, and we adopt metrics using features extracted with this method. Inclusion of these new metrics is inspired by the pioneering theory work for detecting the black hole shadow of Sgr A* presented in \citet{psaltis2015}. As an example, we apply our new technique, metrics and the feature extraction method to data obtained from simulated observations of Sgr A* with the array of the EHT expected in Spring 2017/2018. 

\section{Imaging Methods and the Total Square Variation}\label{sec:2}
As in our previous work \citep{akiyama2017a,akiyama2017b}, we aim to solve for the optimal two-dimensional image reconstruction $\textbf{I} = \{I_{i,j}\}$ in Stokes $I$ by using the following equation \citep[see][for theoretical background]{akiyama2017b}: 
\begin{equation}
    \textbf{I} = \argmin _{\textbf{I}} \left( \chi^2(\textbf{I}) + \Lambda _\ell ||\textbf{I}||_1 + \Lambda _t ||\textbf{I}||_{\textrm{tv}}\right)~\textrm{s.t.}~I_{i,j} \geq 0 \label{eq:imaging}
\end{equation}
where $||\textbf{I}||_p$ is the $\ell _p$ norm of the vector $\textbf{I}$ given by
\begin{equation}
    ||\textbf{I}||_p = \left( \sum _i \sum _j |I_{i,j}|^p \right)^{\frac{1}{p}}\,\,\,\textrm{(for~}p>0\textrm{)}, \label{eq:lp norm}
\end{equation}
and $||\textbf{I}||_\textrm{tv}$ indicates an operator for TV described below.

The first term in Equation (\ref{eq:imaging}) is the traditional $\chi ^2$-term that represents the deviation between the reconstructed image and the observed data. The form of the $\chi ^2$-term depends on the type of data. For measurements of full complex visibilities, it can be given by the residual sum of squares (RSS) between the model and observed full complex visibilities \citep[e.g.][]{honma2014,ikeda2016,akiyama2017b},
\begin{equation}
    \chi^2(\textbf{I}) = ||\textbf{V} - \textbf{F}\textbf{I}||^2_2,
    \label{eq:chisq_fullvis}
\end{equation}
where $\textbf{V}$ is the observed visibility and $\textbf{F}$ is the matrix of a discrete Fourier transform. For measurements of the visibility amplitude $\bar{\textbf{V}}=\{|V_i|\}$ and closure phase ${\bf \Psi}$, a popular form of the $\chi ^2$-term is given by \citep[e.g.][]{buscher1994,lu2012,lu2013,akiyama2017a}
\begin{equation}
    \chi^2(\textbf{I}) = ||\bar{\textbf{V}} - \textbf{A}(\textbf{F}\textbf{I})||^2_2 + ||{\bf \Psi} - \textbf{B}(\textbf{F}\textbf{I})||^2_2, \label{eq:chisq_bispec}
\end{equation}
where \textbf{A} and \textbf{B} indicate operators to calculate the visibility amplitude and closure phase, respectively. 
Following our previous work \citep{akiyama2017a,akiyama2017b}, we normalize deviations between the model and observed data in each $\ell_2$ norm term with the errors of corresponding data, by multiplying deviations with a diagonal matrix ${\bf W}=\{\delta _{i,j} /\sigma _i^2\}$ where $\sigma_i$ is observational error of each data point and $\delta _{i,j}$ is the Kronecker delta. 

The second term in Equation (\ref{eq:imaging}) represents regularization using the $\ell _1$ norm, a sparse regularization function for the image. This is equivalent to the total flux under the non-negative condition assumed in this paper. $\Lambda _\ell $ is the regularization parameter, adjusting the degree of sparsity --- in general, a large $\Lambda _\ell $ leads to a solution with very few non-zero components (i.e., a sparse image), while a small $\Lambda _\ell $ prefers a non-sparse solution.

The third term in Equation (\ref{eq:imaging}) is the TV regularization, defined by the sum of all differences of the brightness between adjacent image pixels. This is a good indicator of image sparsity in its gradient domain instead of the image domain. Minimizing the TV leads to a smooth image that has a small number of pixel-to-pixel brightness variations. As introduced in \S\ref{sec:1}, we adopt two forms of the TV term for the TV regularization.

The first one is isotropic TV \citep{rudin1992} (henceforth isoTV), which has been adopted for astronomical imaging in the literature \citep[e.g.][]{wiaux2010,mcewen2011,carrillo2012,carrillo2014,uemura2015,chael2016}, including our previous work \citep[][]{ikeda2016,akiyama2017a,akiyama2017b}. The isoTV term is a convex function defined by 
\begin{equation}
    ||\textbf{I}||_{\textrm{isotv}}=\sum_{i} \sum_{j} \sqrt{|I_{i+1,j}-I_{i,j}|^2+|I_{i,j+1}-I_{i.j}|^2}. \label{eq:isoTV}
\end{equation}
In general, isoTV regularization leads to a smooth but edge-preserved image, as introduced in \S\ref{sec:1}. 

In this paper, we propose a new form of the TV term, named {\it Total Squared Variation} (henceforth TSV), which is a convex function defined by
\begin{equation}
    ||\textbf{I}||_{\textrm{tsv}}=\sum_{i} \sum_{j} \left( |I_{i+1,j}-I_{i,j}|^2+|I_{i,j+1}-I_{i.j}|^2 \right). \label{eq:TSV}
\end{equation}
Like isoTV, the TSV term takes the sum of all differences of the brightness between adjacent image pixels.  Unlike isoTV, the TSV term does not take the square root of the brightness differences. This slight modification makes the reconstructed image edge-smoothed. For diffuse astronomical images, edge-smoothed images may be a better representation of the truth image, in which case TSV regularization may achieve better performance. An obvious advantage of the TSV term compared with post-processing smoothing (such as convolution with a Gaussian beam, as adopted in CLEAN) is that the TSV term can reconstruct edge-smoothed images while maintaining consistency with observational data.  In contrast, smoothing in post-processing introduces deviations between the smoothed reconstructed image and observational data.

The difference between the isoTV and TSV terms --- edge-preserving or edge-smoothing --- can be qualitatively understood by considering their behavior in response to a brightness offset $\Delta I$ between two adjacent pixels. For a given $\Delta I$, the isoTV and TSV terms will be increased by $\Delta I$ and $\Delta I ^2$, respectively. For small discrepancies, the isoTV term will give a stronger penalty than TSV term.  Larger edgelike differences will be more effectively regularized with TSV term. As a result, TSV regularization will favor a smooth image with smaller variations between adjacent pixels, while isoTV regularization will preserve more edgelike features. 

Equation \ref{eq:imaging} is a convex minimization for full complex visibilities (Equation \ref{eq:chisq_fullvis}) even without the non-negative condition, which gives a unique solution regardless of the initial conditions. Many efficient algorithms have been proposed to solve this problem, such as the fast iterative shrinking threshing algorithm \citep[FISTA;][]{beck2009a, beck2009b}. For instance, the problem with both $\ell _ 1$ and TSV regularizations can be solved with a monotonic variant of FISTA (MFISTA) described in \citet{akiyama2017b}.

On the other hand, when the input data consist of visibility amplitudes and closure phases, the problem becomes non-linear and non-convex.  In this work, we adopt the non-linear programming algorithm L-BFGS-B \citep[][]{byrd1995,zhu1997} to solve the equation, as in our previous work \citep{akiyama2017a}. Because the problem becomes non-convex for this type of data, a global solution is generally not guaranteed, which is common with other imaging techniques using closure quantities \citep[e.g.][]{buscher1994,thiebaut2008,bouman2016,chael2016,akiyama2017a}. 

\section{Simulated Observations and Imaging}\label{sec:3}
\subsection{Models}\label{sec:3.1}
In this work, we adopt two types of models to examine the performance of imaging techniques presented in $\S$\ref{sec:2}. These models are located at the position of Sgr A*, and synthetically observed with the EHT (see \S\ref{sec:3.2}). 

First, we adopt simple geometric models consisting of two point sources with a flux of 1~Jy to test the effective angular resolution of imaging techniques along with its literal definition --- a minimal angular scale that can identify two sources separately. Because the synthesized beam (henceforth the CLEAN beam) of the EHT has a position angle of $85.5^\circ$ (see \S\ref{sec:3.2}) close to the Right Ascension (RA) axis, we simulate two point sources located along the RA direction.  We examine ten models with different separations ranging from $\delta\theta=0.1\theta_{\rm maj}$ to $1.0\theta_{\rm maj}$, where $\theta_{\rm maj}=22.7$~${\rm \mu}$as is the major axis size of the CLEAN beam.


\begin{figure}
\centering
\includegraphics[width=0.8\columnwidth]{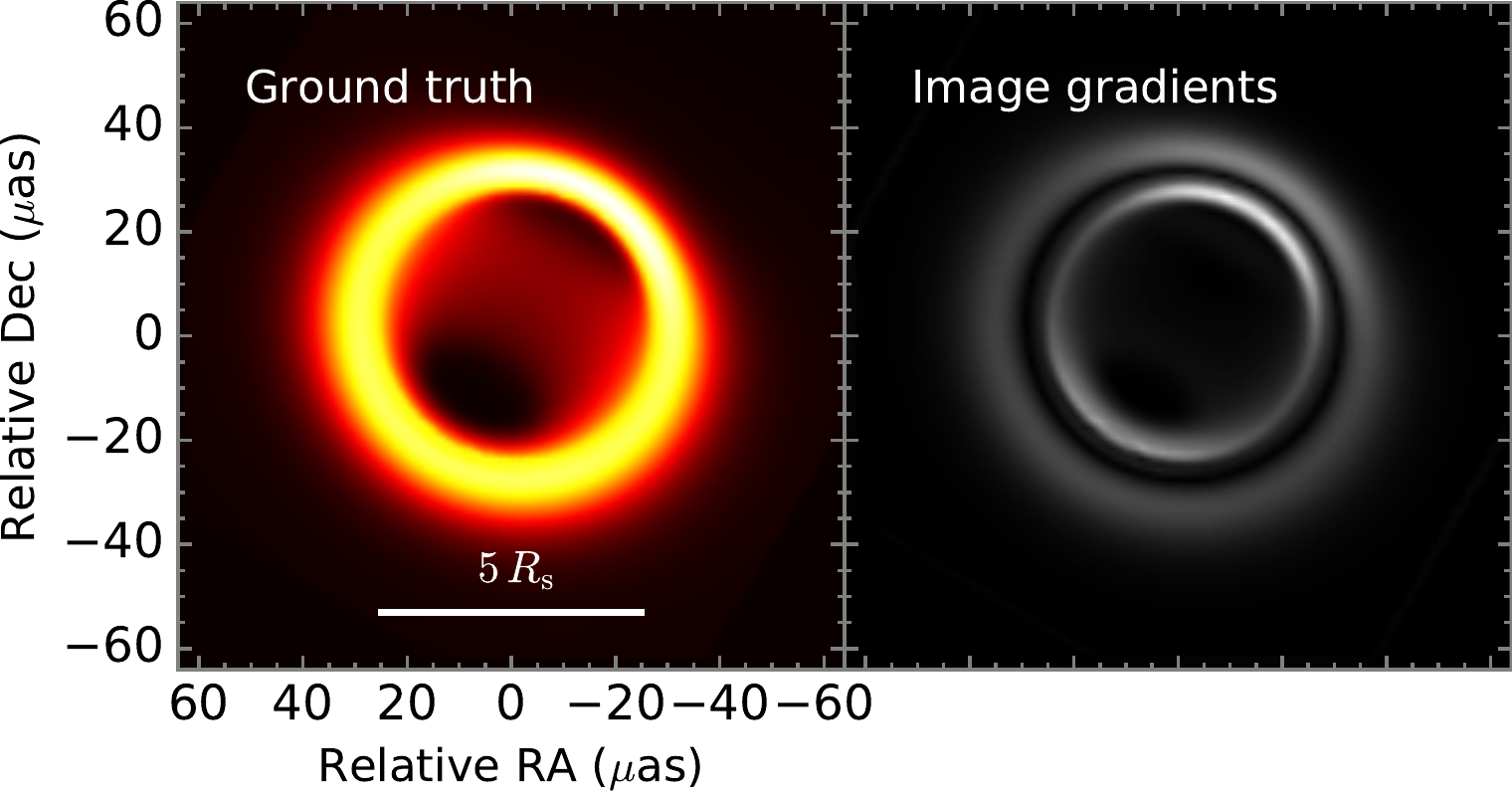} \\
(a) Free-fall Model \vspace{0.5em}\\
\includegraphics[width=0.8\columnwidth]{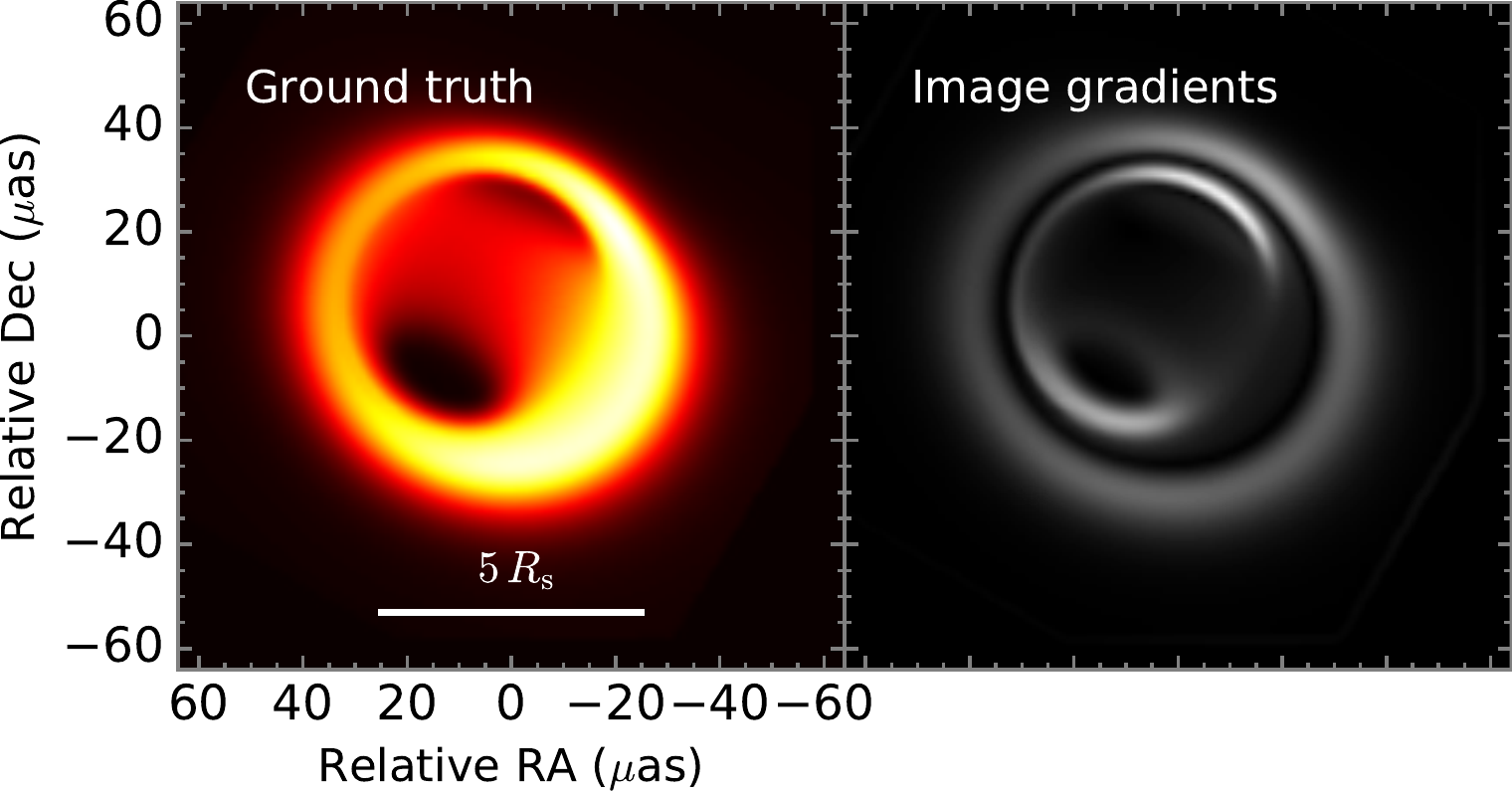} \\
(b) Sub-Keplerian Model \vspace{0.5em}\\
\includegraphics[width=0.8\columnwidth]{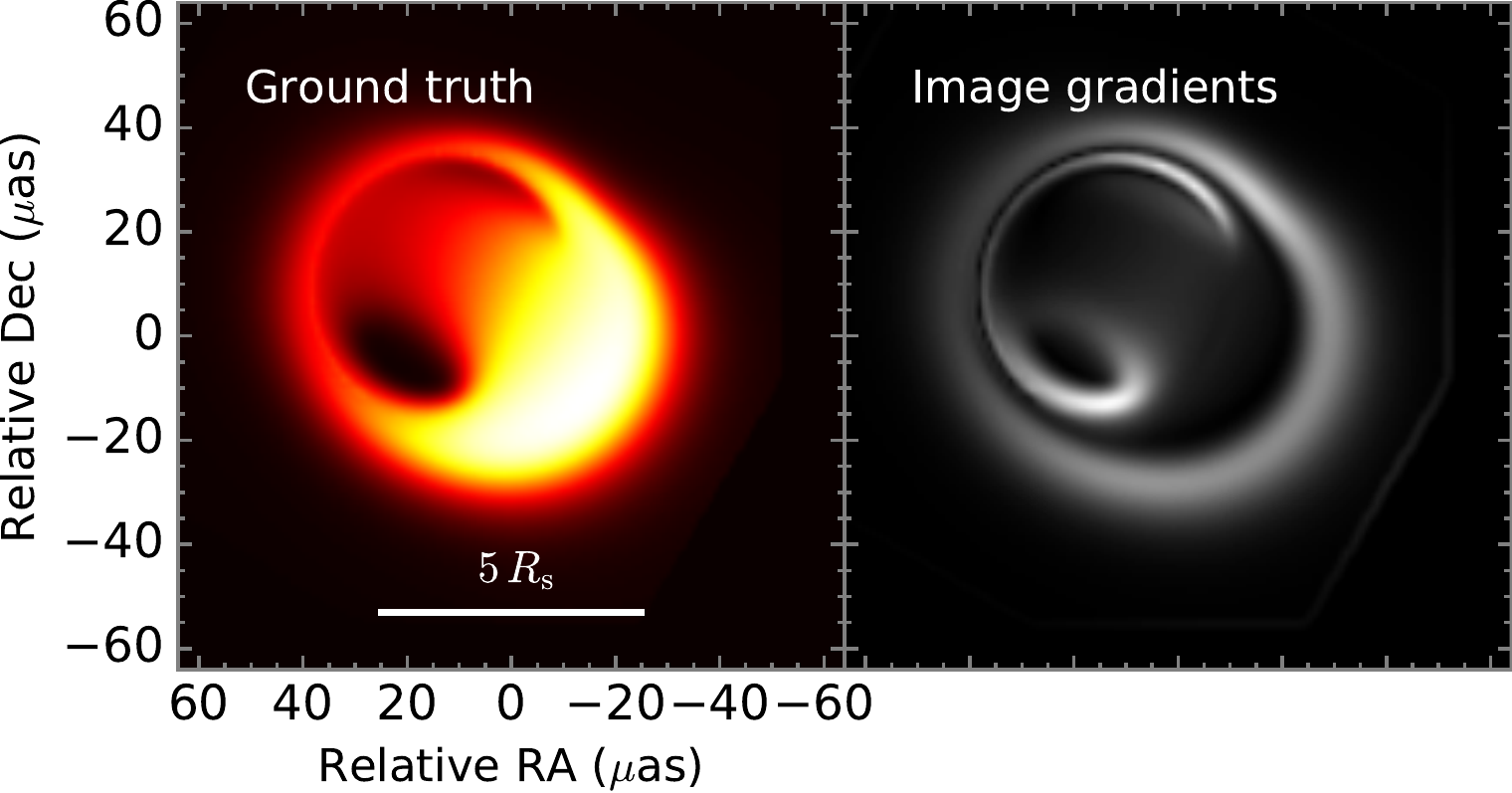} \\
(c) Keplerian Model \vspace{0.5em}
\caption{Three semi-analytic RIAF models around a non-spinning black hole with different toroidal velocities (Free fall, Sub-Keplerian and Keplerian) \citep[see \S\ref{sec:3.1} and ][]{pu2016}. The left panel is the original model image. The right panel is the image gradient of the model image, enhancing the edges of the image (see \S\ref{sec:4.1} for details).
}
\label{fig:model_images}
\end{figure}

Second, we use semi-analytic models of a radiative inefficient accretion flow (RIAF) around a non-spinning black hole in Sgr A$^{*}$ presented in \cite{pu2016} to evaluate the performance of imaging techniques on more complicated and physical images. We adopted three representative images at 1.3~mm (230~GHz) for flow models with dynamics characterized by (i) a Keplerian shell that is rigidly rotating outside the innermost stable circular orbit (ISCO) and in-falling with a constant angular momentum inside the ISCO (henceforth Keplerian Model), (ii) Sub-Keplerian motion (henceforth Sub-Keplerian Model), and (iii) free-falling motion with zero angular momentum at infinity (henceforth Free-fall Model). All three images, shown in Figure~\ref{fig:model_images}, are broadly consistent with visibility amplitude measurements of prior EHT observations in 2009 and/or 2013 \cite[see][for details]{pu2016}.

\subsection{Simulated Observations}\label{sec:3.2}
\begin{table}[t]
	\caption{Stations in simulated observations}
	\centering
	\begin{tabular}{l r} \hline \hline
		Telescope & SEFD (Jy) \\ \hline
		Phased ALMA & 100 \\
		Phased SMA and JCMT & 3600 \\
		LMT & 1400 \\
		IRAM 30m & 1400 \\
		NOEMA single dish & 5200 \\
		ARO/SMT & 11000 \\
		SPT & 9000 \\ \hline
	\end{tabular}
	\label{tb:3-array}
\end{table}

\begin{figure}
	\centering
	\includegraphics[width=1.0\columnwidth]{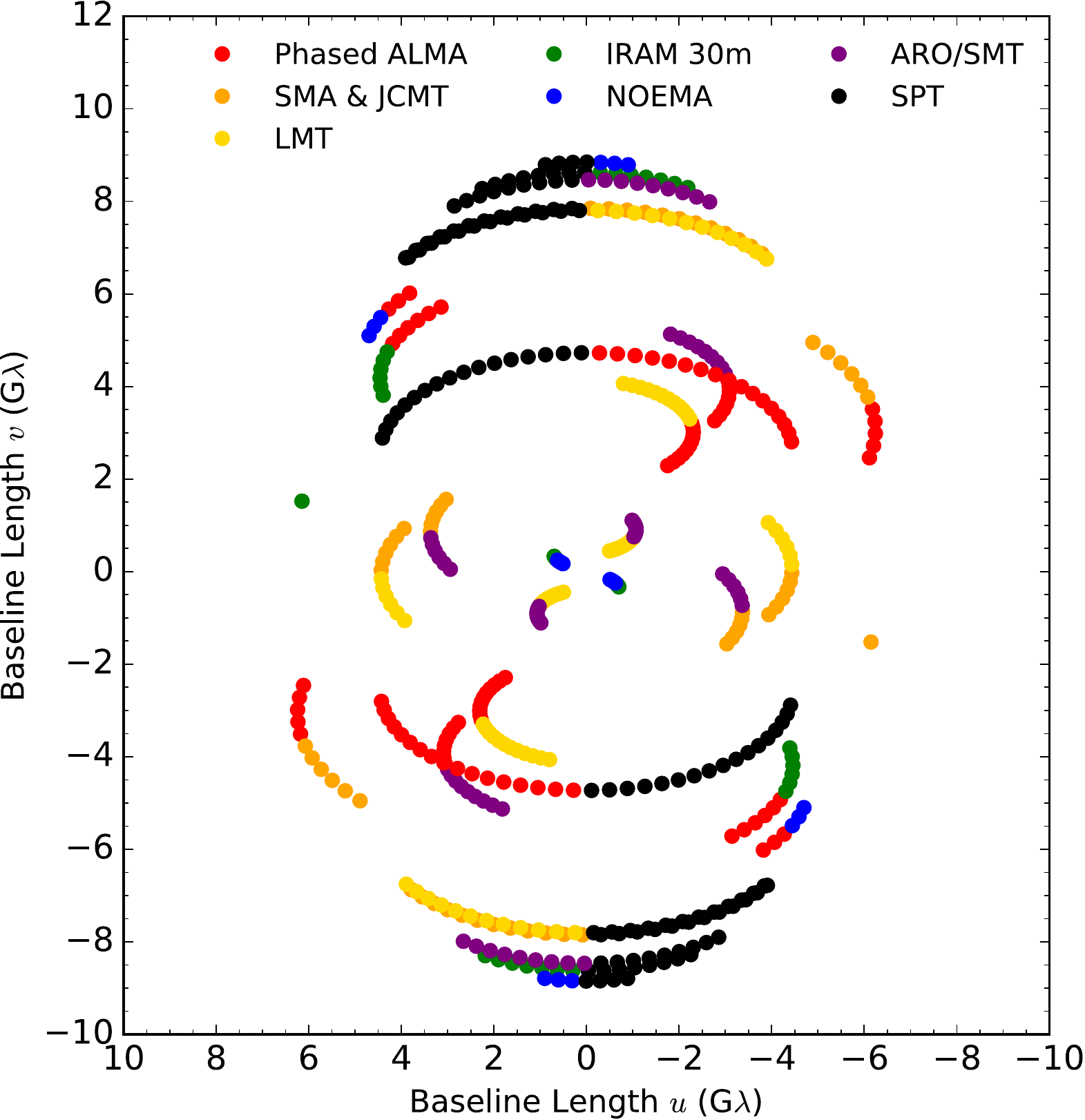}
	\caption{The $uv$-coverage of the simulated EHT observations. Each track is split into two colors to identify the pair of stations in the baseline.}
	\label{fig:uvcoverage}
\end{figure}

We simulate observations of the three Sgr A* models described in \S\ref{sec:3.1} with the EHT at a wavelength of 1.3~mm (230~GHz) using the MIT Array Performance Simulator (MAPS)\footnote{http://www.haystack.mit.edu/ast/arrays/maps/}. The seven stations used in the simulations are listed in Table \ref{tb:3-array} with their expected system equivalent flux densities (SEFD) for observations in 2017/2018. The simulated observations contain a series of 6-minute scans every 20 minutes over a day, and data are integrated for each scan. Synthetic data are generated in Stokes $I$ with a bandwidth of 8 GHz, equivalent to 4 GHz bandwidth in each of two circular polarizations. We adopt other observation parameters as in \cite{akiyama2017a, akiyama2017b}. Figure~\ref{fig:uvcoverage} shows the $uv$-coverage of the simulated observations. The maximum baseline length is 9.0~G$\lambda$, corresponding to $\theta=\lambda/D_{\rm max}=22.9$~${\rm \mu}$as. The major and minor axes and position angle of the CLEAN beam at natural weighting is $22.7$~${\rm \mu}$as, $11.2$~${\rm \mu}$as and $85.5^\circ$, respectively.

For physically motivated models (\S\ref{sec:3.1}), we convolve the physical model images with an elliptical Gaussian based on the scattering law of Sgr A* measured in \cite{bower2006} prior to simulating the observation, which represents angular broadening effects due to diffractive interstellar scattering \citep[e.g.][]{goodman1989,narayan1989,johnson2015a,johnson2016b}. The angular broadening effect is invertible, and results in a net increase in the noise levels of visibilities inversely proportional to the scattering kernel function \citep[e.g., see][for details]{fish2014}. Along with \cite{fish2014}, un-scattered (i.e.~intrinsic) visibility estimates are derived by dividing observed visibilities with the scattering kernel function. Note that, in this paper, we do not include the effects of refractive interstellar scattering in our synthetic observations, which introduce compact refractive substructure into the observed image \citep{johnson2015a,johnson2016a,johnson2016b,akiyama2016b}. The effects of the refractive scattering on the image are \emph{not} simply invertible --- the mitigation of refractive scattering is an ill-posed problem that requires additional regularization \citep[scattering optics;][]{johnson2016c}. Since the scope of this work is testing TSV regularization in interferometric imaging, inclusion of the scattering optics is not essential. We will include scattering optics in our imaging algorithms in future work. We note that here intra-day time variabilities in the $R_s$-scale structure expected for Sgr A* are ignored, since we here focus on evaluation of static imaging. We will discuss issues related to time variability in \S\ref{sec:5}.

\subsection{Imaging}\label{sec:3.3}
We reconstruct images of two point source models and physical models using techniques with $\ell_1+{\rm isoTV}$ and $\ell_1+{\rm TSV}$ regularizations described in \S\ref{sec:2}. Two-point source models are reconstructed using full-complex visibilities, while RIAF models are reconstructed with visibility amplitudes and closure phases. As in \citet{akiyama2017a,akiyama2017b}, we use normalized regularization parameters defined as
\begin{eqnarray}
&&\tilde{\Lambda}_{\rm \ell} \equiv \Lambda_{\rm \ell} \frac{\max ({\bf \bar{V}})}{N_{\rm amp}+N_{\rm cphase}},\\
&&\tilde{\Lambda}_{\rm isotv} \equiv \Lambda_{\rm \ell} \frac{\max ({\bf \bar{V}})}{4(N_{\rm amp}+N_{\rm cphase})},\\
&&\tilde{\Lambda}_{\rm tsv} \equiv \Lambda_{\rm \ell} \frac{\max ({\bf \bar{V}}^2)}{4(N_{\rm amp}+N_{\rm cphase})}.
\end{eqnarray}
which are less affected, compared to unnormalized regularization parameters, by the total flux density of the target source ($\approx \max {\bf \bar{V}}$) and the number of data points in visibility amplitudes ($N_{\rm amp}$) and closure phases ($N_{\rm cphase}$). We adopt $\tilde{\Lambda}_{\rm \ell}=10^3,10^2,10^1,10^0$, $\tilde{\Lambda}_{\rm isotv}=10^4,10^3,10^2,10^1$, and $\tilde{\Lambda}_{\rm tsv}=10^4,10^3,10^2,10^1$, providing 16 parameter sets for both $\ell_1+{\rm isoTV}$ and $\ell_1+{\rm TSV}$ regularizations. The optimal imaging parameters are determined from these parameter sets with 10-fold cross validation \citep[see][for details]{akiyama2017a}.

In addition, we reconstruct images with the Cotton-Schwab CLEAN algorithm \citep[henceforth, CS-CLEAN;][]{schwab1984} implemented in the Common Astronomy Software Applications (CASA) package\footnote{\url{https://casa.nrao.edu/}}. We adopt a field of view (FOV) of 130~$\mu$as, gridded into 128~pixels in each of right ascension and declination. The resulting pixel size of $\sim 1.02$~${\rm \mu}$as corresponds to a physical scale of $\sim 0.1$~$R_s$. 
In general, the performance of CLEAN algorithms is highly affected by the choice of $uv$-weighting and CLEAN gains \citep[e.g.][]{thompson2017}. We adopt different $uv$-weighting schemes (uniform, natural, Briggs with robust parameters of $\{-2,-1,0,1,2\}$) and also CLEAN gains (0.1, 0.5, 1, 5, 10\%). In this paper, we show results at uniform weighting and CLEAN gains of 10~\%, stably showing the best performance for all three models. Note that this relatively high value of optimal CLEAN gains would be due to the source structure of the adopted physical models, which have a complicated structure on scales comparable to the CLEAN beam size.

\section{Evaluation of the Image Fidelity}\label{sec:4}
We evaluate the performance of the image reconstruction using image fidelity metrics as in previous literature. In this work, we adopt several new metrics that evaluate similarities in other types of features. In addition, as in other recent VLBI imaging work \citep{chael2016, akiyama2017a, akiyama2017b}, we adopt the NRMSE metric on the image, defined by
\begin{equation}
    \textrm{NRMSE}_{\textrm{image}}(\textbf{I},\textbf{K}) = \sqrt{\frac{\sum_{i}\sum_{j}|I_{i,j}-K_{i,j}|^2}{\sum_{i}\sum_{j} |K_{i,j}|^2}},
    \label{eq:NRMSE}
\end{equation}
where $\textbf{I}=\{I_{i,j}\}$ and $\textbf{K}=\{K_{i,j}\}$ are the image to be evaluated and the reference image, respectively. Since larger deviations in the evaluated image from the reference image will provide larger NRMSEs, images with smaller NRMSEs can be interpreted with good imaging fidelities. 

\subsection{A Metric Using the Image Gradient}\label{sec:4.1}
One of the physically most important features in the images is the width or size of the emission region, or more generally how and where the brightness distribution varies, which is useful to get the outline of the brightness distribution. Such spatial variations in the brightness distribution can be enhanced by taking the gradient of the image \citep[e.g.][]{psaltis2015}, which is given by
\begin{equation}
  |\nabla I(x,y)| = \sqrt{\left| \frac{\partial I}{\partial x} \right |^2+\left| \frac{\partial I}{\partial y} \right|^2}
    \label{eq:gradient}
\end{equation}
for a continuous brightness distribution. There are many methods to approximate the derivatives in Equation (\ref{eq:gradient}) for discrete images \citep[e.g.~see][for a review]{cui2013}. A popular method is the Sobel approximation, given by
\begin{eqnarray}
\frac{\partial I_{i,j}}{\partial x} \approx I_{i+1,j-1} + 2I_{i+1,j} + I_{i+1,j+1} \nonumber \\
- I_{i-1,j-1} - 2I_{i-1,j} - I_{i-1,j+1}. \label{eq:sobel_x}\\
\frac{\partial I_{i,j}}{\partial y} \approx I_{i-1,j+1} + 2I_{i,j+1} + I_{i+1j+1} \nonumber \\
- I_{i-1,j-1} - 2I_{i,j-1} - I_{i+1,j-1}, \label{eq:sobel_y}
\end{eqnarray}
The Sobel approximation takes the central differences of an image pixel where the gradient is estimated and its four adjacent pixels, giving double weight to the central pixel. In this work, we adopt the Sobel approximation to calculate the image gradient. We use the python implementation in {\tt scikit-image}\footnote{http://scikit-image.org/\label{fnt:scikit-image}}. We show gradients of the three model images in the middle panels of Fig \ref{fig:model_images}. Errors in the image gradients can be evaluated by taking the NRMSE for them,
\begin{equation}
    \textrm{NRMSE}_{\textrm{grad}} (\textbf{I},\textbf{K}) \equiv \textrm{NRMSE}_{\textrm{image}} (\nabla \textbf{I},\nabla \textbf{K}).
\end{equation}
This metric is more sensitive to steep gradients in the brightness distribution, namely edges of the emission structure, than to their brightness.

\subsection{Metrics Using the Circle Hough Transform}\label{sec:4.2}
For astronomical objects with ring-like morphology, such as black hole shadows, accretion disks and cavity-like structures, the radius and central position of its curvature are often useful for extracting physical information.  In the case of the black hole shadow (i.e., the photon sphere of the black hole), its apparent shape and size depend on the mass and spin of the black hole and the viewing orientation of the observer. Its apparent diameter ranges from $\sqrt{27} \sim 5.2$~$R_s$ for a non-rotating black hole \citep[independent of viewing angle;][]{bardeen1973} to $4.84$~$R_s$ for a maximally spinning one viewed pole-on \citep[see][and references therein]{psaltis2015}. As a result, its apparent diameter changes only $\sim 4$~\% with black-hole spin and viewing orientation. Measuring its radius provides a unique opportunity to test general relativity \citep[][]{psaltis2015}, and provide independent measurements of the mass-to-distance ratio of the super-massive black holes.

In this work, we adopt new metrics using the circle Hough transform \citep[CHT; e.g.][]{duda1972}, which is a subclass of the Hough transform \citep[e.g.][]{hough1962} that is better suited to detecting circular structures in an image. We introduce the CHT in \ref{sec:4.2.1} and metrics using the CHT in \ref{sec:4.2.2}. Other approaches presented in \citet{psaltis2015} that utilize the linear Radon transform \citep[][]{radon1986} and/or the linear Hough transform \citep[e.g.][]{hough1962,duda1972} to derive the radius and center of the emission illuminating the photon sphere. These techniques can be generalized to extract any features. The CHT presented in this work is specialized and simplified to detect circles in an image. Since the shape of the black hole shadow is nearly circular, we take the simple approach of extracting the radius of the photon ring with the CHT. 

\subsubsection{Circle Hough Transform}\label{sec:4.2.1}
The circle Hough transform (CHT) is a feature extraction method specialized for circle detection. The most classical form of the CHT \citep[e.g.][]{duda1972} derives a three-dimensional accumulator from the input two-dimensional image $I(x,y)$, as given by
\begin{eqnarray}
  &&H(x,y,r;I)=\iint _{C} I(x',y')\,dx'dy',~~~\textrm{where}\nonumber\\
  &&C=\{(x',y');\,(x'-x)^2+(y'-y)^2=r^2\}.
  \label{eq:hough_accumulator1}
\end{eqnarray}
At a position of the image $(x,y)$, the classic CHT integrates the brightness distribution on a circle with the radius $r$ centered there. The accumulator will be enhanced if a circular feature with the corresponding radius is located there. Therefore, the magnitude of the accumulator is a good indicator of circles in the image, and can be considered as a probabilistic description for the existence of circles. 

An important advantage compared with the linear Hough/Radon transforms presented in previous work \citep[][]{psaltis2015} is that the distribution of the accumulator is not affected by the choice of the reference (i.e.~origin) position. Linear Hough/Radon transforms are highly affected by the reference position, and generally require a four-dimensional search space (i.e., radius, position angle and the location of the origin) to detect circles. Therefore, the CHT provides more mathematically and intuitively straightforward indicator of circles in the image.

In this work, we adopt a discretized form of Eq. (\ref{eq:hough_accumulator1}) given by
\begin{eqnarray}
  &&H(x,y,r;I)= \sum_{i=1}^{N_p} I(x-\Delta x_i,y-\Delta y_i),\nonumber\\
  &&\textrm{where}\,\,\,\Delta x_i = r \cos \frac{2\pi i}{N_p},\,\,\,\Delta y_i=r\sin \frac{2\pi i}{N_p}.
  \label{eq:hough_accumulator2}
\end{eqnarray}
In Eq. (\ref{eq:hough_accumulator2}), the CHT is discretized in the position-angle direction, and we adopt $N_p=360$, which provides a resolution of 1$^\circ$ for the position angle. Since the shifted coordinate $(x-\Delta x_i,y-\Delta y_i)$ is mostly off-grid, the brightness on each coordinate is evaluated with a bi-cubic spline interpolation. We have also examined larger values for $N_p$ (e.g. $N_p=3600$) and confirm that the quantities derived from the CHT are consistent and well-converged.


\begin{figure*}
\centering
\includegraphics[width=1.0\textwidth]{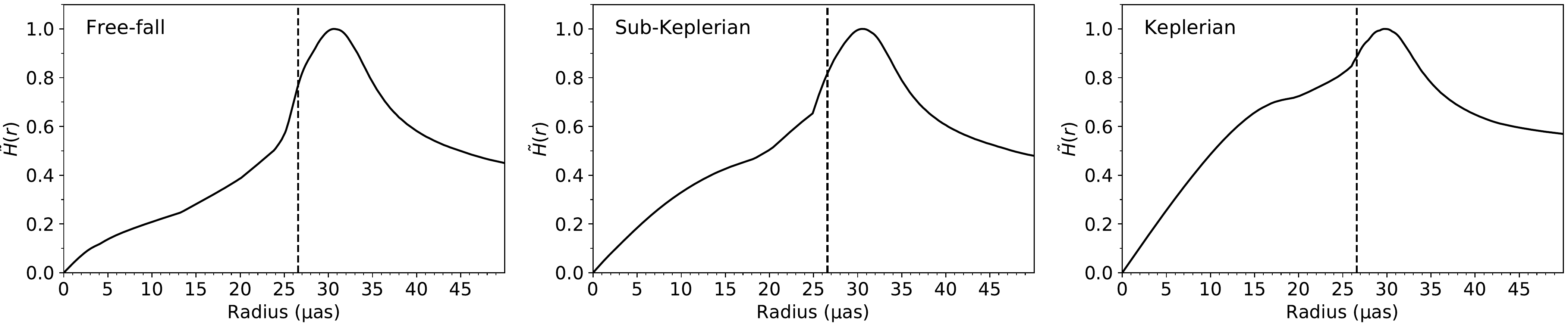} \\
\caption{The radial profile of the maximum CHT accumulator for the three model images.
Each profile is normalized by its maximum value. The dashed lines represent the radius of the black hole shadow for the non-spinning black hole \citep[$\sim 2.6~R_s \sim 27~{\rm \mu as}$;][]{bardeen1973} assumed in the models.}
\label{fig:rad_profile}
\end{figure*}

In Figure~\ref{fig:rad_profile}, we show the radial profile of the maximum CHT accumulator, defined by
\begin{equation}
\tilde{H}(r; I) \equiv \max _{x,y} H(x,y,r; I),
\end{equation}
for three model images. When the CHT accumulator is applied to the unfiltered original image, the principal component with the highest value is at a radius of $\sim 30$ $\rm \mu$as for all three models, regardless of the toroidal velocity. 

Although the peak radii of the ring-like emission are slightly larger than the radius of the black hole shadow　for Sgr A* ($\sim 27$~$\rm \mu$as), they are less affected by the toroidal velocity and also the black hole spin that create highly asymmetric brightness distributions. This indicates that the CHT is an attractive choice to extract the size of the photon ring illuminating the black hole shadow. We show circular features corresponding to representative peaks obtained from the three model images in \S\ref{sec:5.3} with detailed comparison to those obtained from simulated observations.


\subsubsection{Metrics Using Representative Circular Features}\label{sec:4.2.2}
From a feature extraction perspective, circular features derived from the peak in the radial CHT profile $\tilde{H}(r; I)$ are good indicators of the image fidelity. In this paper, we adopt two metrics using circular features derived from the CHT accumulators, which are unaffected by the location of the origin (i.e., the phase-tracking center). These are particularly useful for high-frequency VLBI and optical-interferometric images where the absolute position is not measurable without phase referencing. 

The radii $\rm R_{\rm peak} ({\bf I})$ of the peak in the radial CHT profile are important for estimating the size of the black hole shadow. Here, we define a metric using the peak radii for a corresponding feature in a pair of images, given by
\begin{equation}
    \Delta R ({\bf I},\,{\bf K}) \equiv |\rm R_{\rm peak}({\bf I}) -R_{\rm peak}({\bf K})|.
    \label{eq:radii_offset}
\end{equation}

Another useful quantity is the positional offset ${\bf P}(r_i;{\bf I})$ between the curvature center and the center of the mass of the image. This quantity is independent of the location of the origin and also reflects the asymmetry of the brightness distribution --- one can expect a larger offset for more asymmetric images (i.e., faster toroidal velocity for models adopted in this work). Here, we adopt a metric for the difference in this positional offset between a pair of images defined as
\begin{equation}
    \Delta P ({\bf I},\,{\bf K}) \equiv |{\bf P}(r_i;{\bf I})-{\bf P}(r_i;{\bf K})|.
    \label{eq:pos_offset}
\end{equation}

\section{Results}\label{sec:5}
\subsection{Effective Angular Resolutions}\label{sec:5.1}
\begin{figure*}
\centering
\includegraphics[height=0.9\textheight]{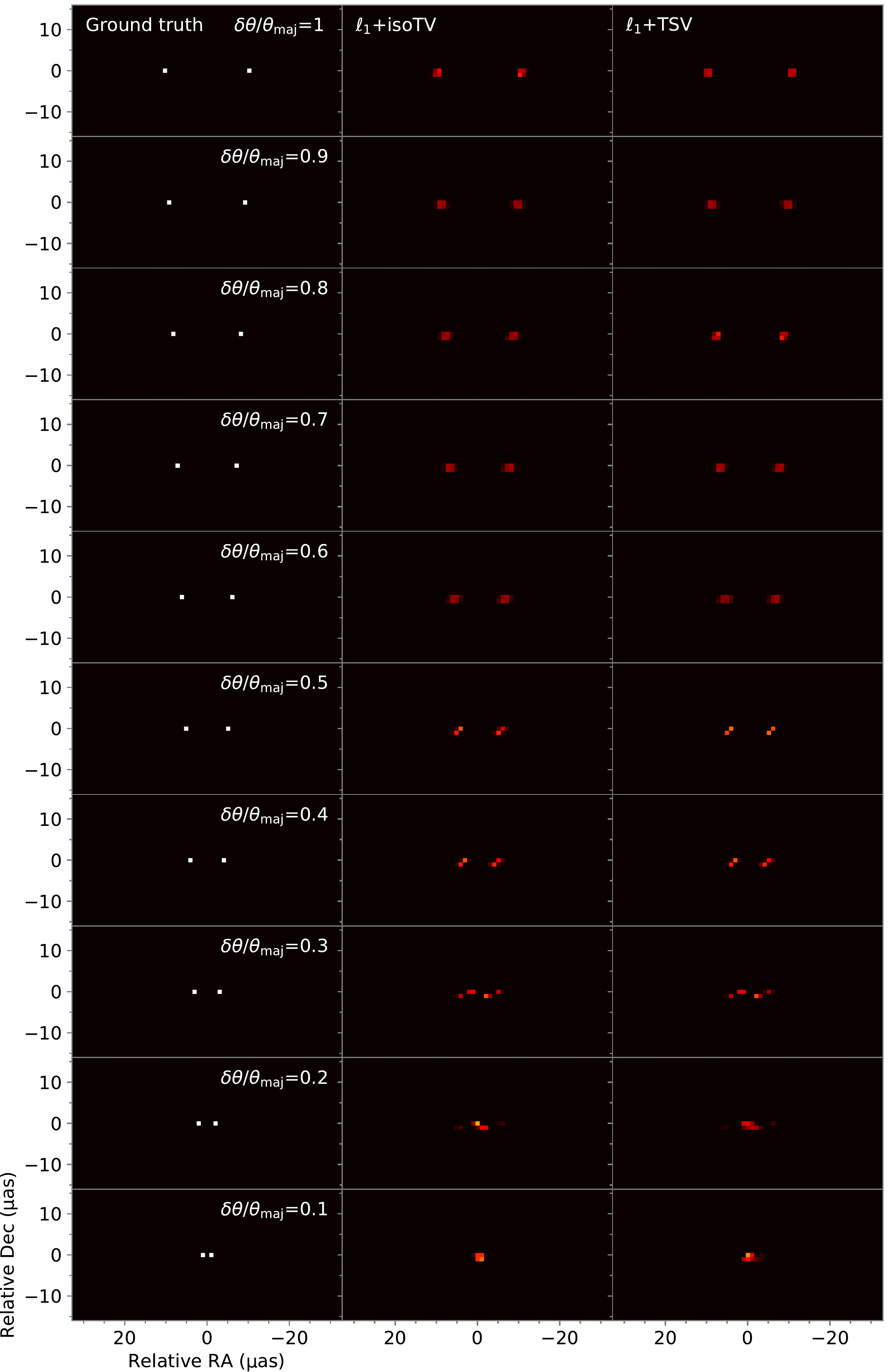}
\caption{The ground truth image (leftmost) and images reconstructed with $\ell_1$+isoTV (middle) and $\ell_1$+TSV (rightmost) regularizations. The distance between two point sources $\delta\theta/\theta _{\rm maj}$ (where $\theta _{\rm maj}=22.7$ $\rm \mu$as) changes from 0.1 to 1 at intervals of 0.1.}
\label{fig:tpsimages}
\end{figure*}
We show the ground-truth and reconstructed images of two point source models in Figure~\ref{fig:tpsimages}.
The reconstructed images are broadly consistent between $\ell_1$+isoTV and $\ell_1$+TSV regularizations. Regardless of separation and regularization, each point source is blurred into a few pixels due to observational noise. Two point sources are clearly separated at separations larger than 0.4 $\theta_{\rm maj}$, while two point sources start to be identified as a single source at separations smaller than 0.3 $\theta_{\rm maj}$. This suggests that superresolution of $\sim$30~\% can be achieved in synthetic observations with these imaging techniques.  

The above results are consistent with previous work, which reconstruct images by directly solving observational equations with convex regularization function(s). For instance, a superresolution of $0.25\lambda/D$ has often been quoted for the Maximum Entropy Method \citep[MEM; see][and reference therein]{narayan1986}, which can be worse depending on the quality of data \citep{holdaway1990}. A similar factor of $\sim 0.25$ is obtained, for instance, in one-dimensional simulations in \citet{honma2014} using $\ell_1$ regularization with observational noise. This superresolution factor of $\sim$0.25 would be the best case which can be obtained in high-quality data with high signal-to-noise ratio and $uv$-coverages, and can be worse if the data have a small signal-to-noise ratio or a poor $uv$-coverage.

Our simulations, which have larger noise and sparser two-dimensional $uv$-coverage,  imply a similar factor of $\sim 30$\%. This is slightly worse than the above work,  but yet broadly consistent. Our results, combined with the previous work, suggest that the regularized imaging techniques can achieve superresolution around $\sim 30$\% for moderate or high SNR data regardless of regularization function, and consistently outperforming greedy image-domain deconvolution algorithms like CLEAN \citep[e.g.][]{hogbom1974,clark1980,schwab1984} and classical MEM techniques \citep{cornwell1985}.

\subsection{Image Appearance of Physical models}\label{sec:5.2}
\begin{figure*}
\centering
\includegraphics[width=0.8\textwidth]{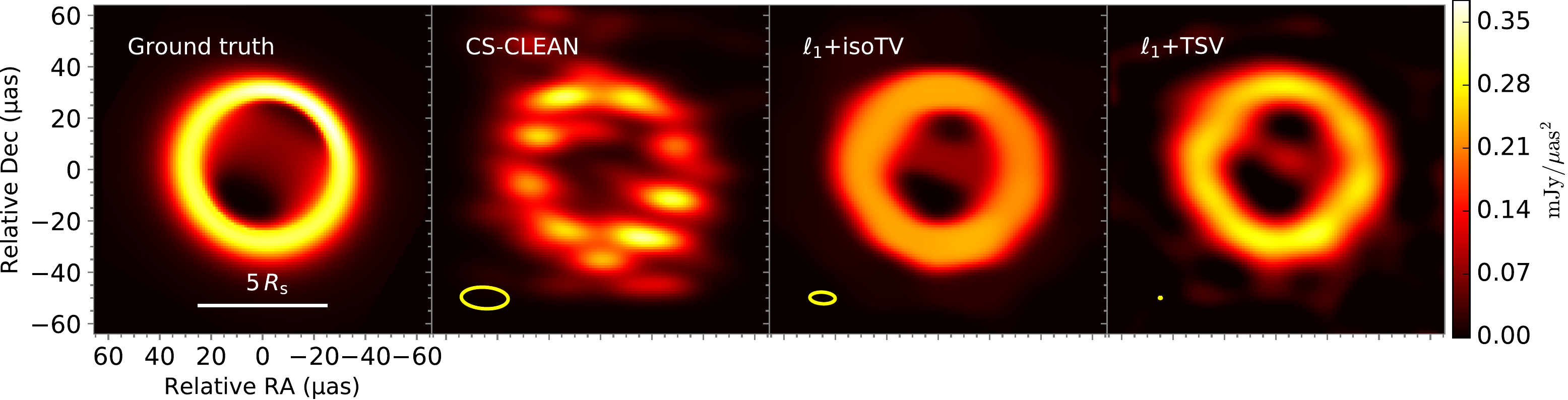} \\
(a) Free-fall Model\vspace{0.5em}\\
\includegraphics[width=0.8\textwidth]{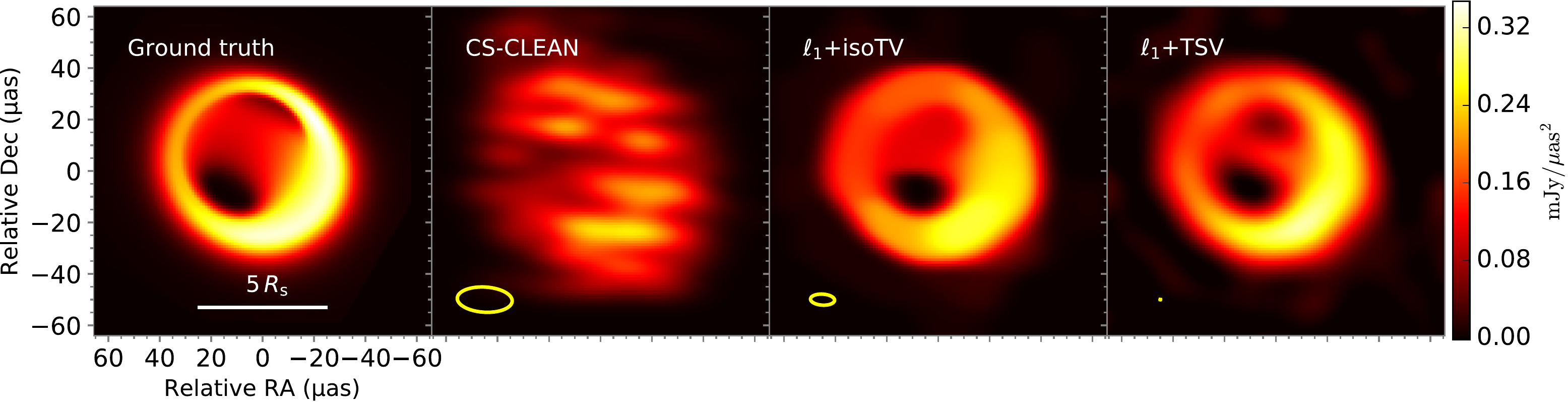} \\
(b) Sub-Keplerian Model\vspace{0.5em}\\
\includegraphics[width=0.8\textwidth]{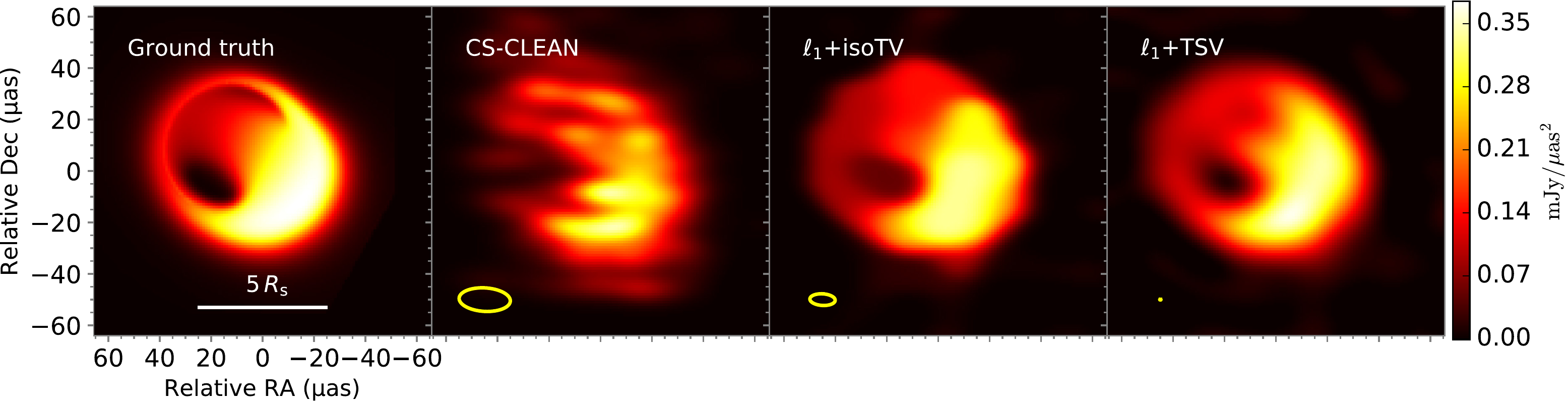}\\
(c) Keplerian Model\\
\caption{The ground truth image (leftmost) and images reconstructed with CS-CLEAN (second from left), $\ell_1$+isoTV (second from right) and $\ell_1$+TSV (rightmost) regularization. All reconstructed images are convolved with elliptical Gaussian beams represented by the yellow ellipses, for which the size corresponds to the optimal resolution determined with the image-domain NRMSE curve in Figure~\ref{fig:nrmses} (see \S\ref{sec:5.3}). The same transfer function is adopted for four images of each model (i.e. on each row). }
\label{fig:opt_recimages}
\end{figure*}

We show the ground-truth images and the images reconstructed with CS-CLEAN, $\ell_1$+isoTV and $\ell_1$+TSV regularizations in Figure~\ref{fig:opt_recimages}. The CS-CLEAN reconstruction uses full-complex visibilities, while the sparse modeling techniques ($\ell_1$+isoTV/TSV) use only visibility amplitudes and closure phases. These images are convolved with restoring elliptical Gaussian beams that minimize the image-domain NRMSE between the unconvolved ground-truth image and beam-convolved reconstructed images (see \S\ref{sec:5.3} for details).

A clear difference between CS-CLEAN and sparse modeling techniques ($\ell_1$+isoTV/TSV) is that CS-CLEAN reconstructs images that are too sparse with many compact artifacts. As explained in \citet{akiyama2017b}, the most critical underlying assumption of CS-CLEAN (and also a pure $\ell _1$ regularization on the image), which is to pursue sparsity in the image, does not work well by itself, a result that is consistent with previous work \citep[e.g.][]{chael2016, akiyama2017a}. In the Keplerian model in particular, the CS-CLEAN image is too sparse to reconstruct the faint ring-like emission on the east side of the image. On the other hand, $\ell_1$+isoTV/TSV regularizations reconstruct much smoother images with ring-like emission for all three ground truth images. CS-CLEAN also incorrectly puts point-source artifacts in a more extended region than the ground truth image and $\ell_1$+isoTV/TSV reconstructions. 

\begin{figure*}
\centering
\includegraphics[width=0.7\textwidth]{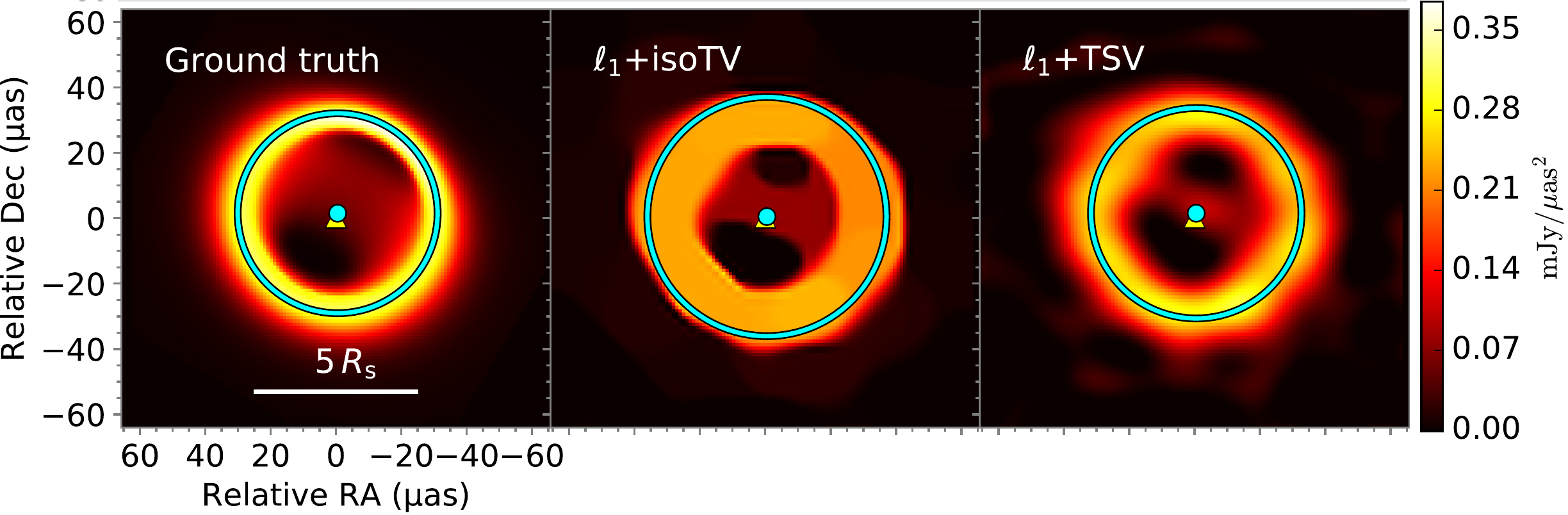} \\
(a) Free-fall Model\vspace{0.5em}\\
\includegraphics[width=0.7\textwidth]{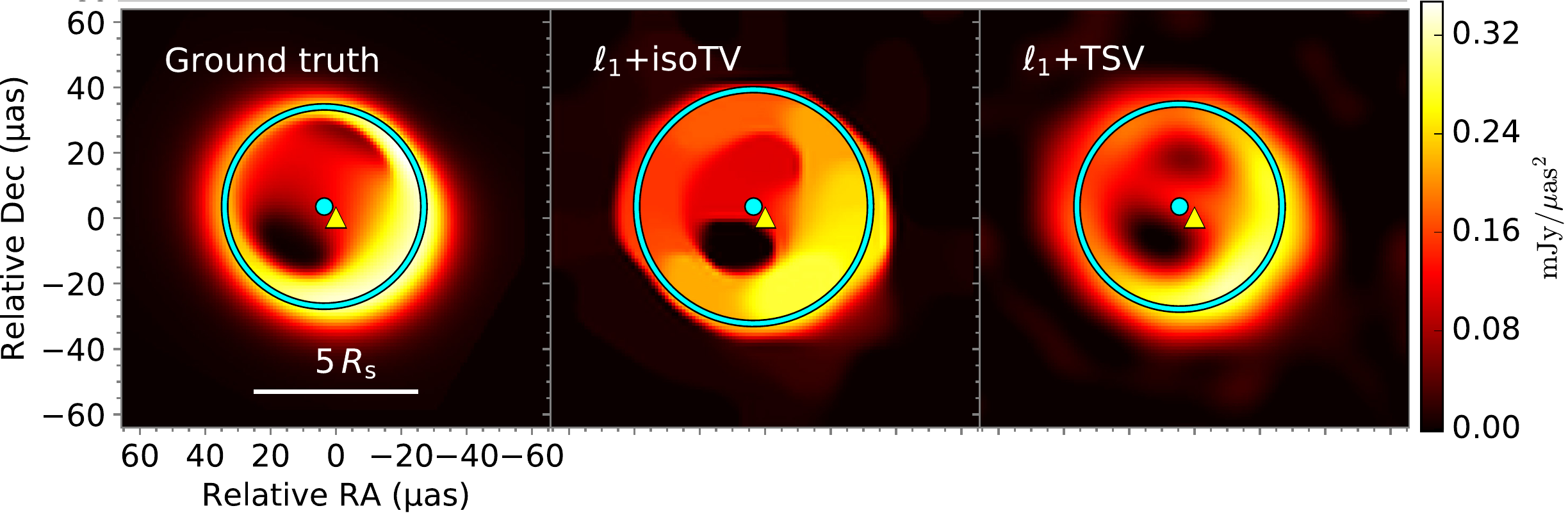} \\
(b) Sub-Keplerian Model\vspace{0.5em}\\
\includegraphics[width=0.7\textwidth]{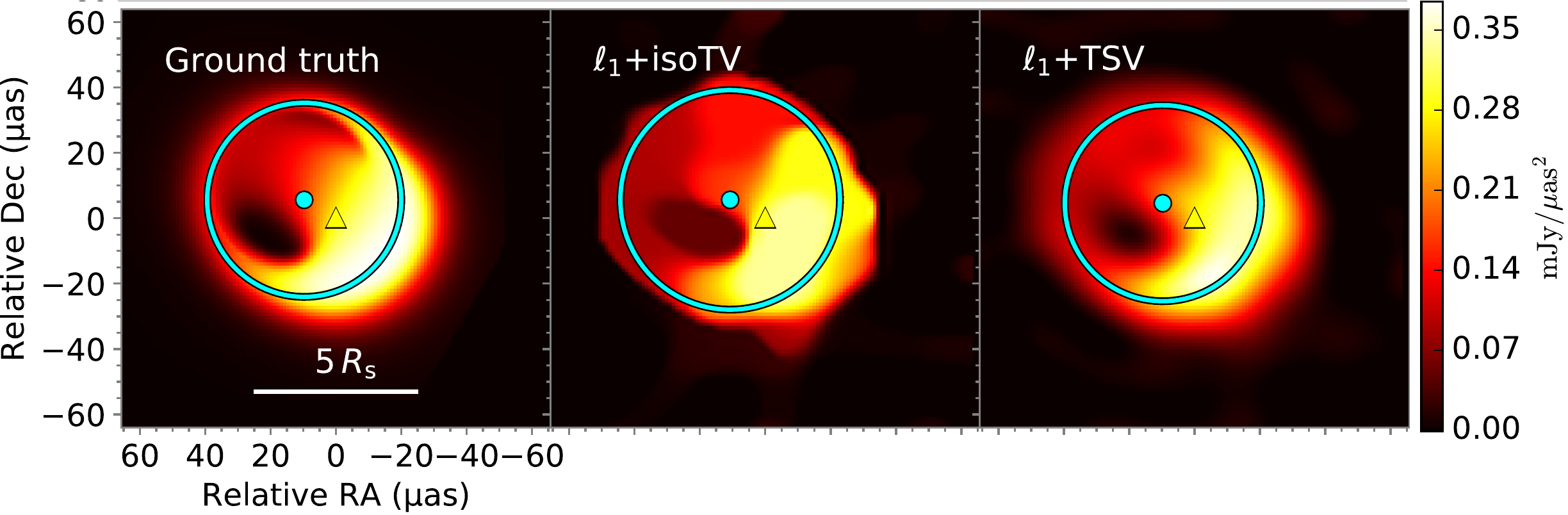}\\
(c) Keplerian Model\\
\caption{The ground truth and unconvolved (raw) reconstructed images with $\ell_1$+isoTV/TSV regularizations. We overlay circular features detected with the CHT. Colored dots indicate the central positions of the correspondingly colored circles derived from the CHT accumulator (see \S\ref{sec:5.4} for details). Filled triangles represent the centers of the mass of the images. Blue circles represent the principal circular feature in the CHT accumulator of the images. 
}
\label{fig:raw_recimages}
\end{figure*}

Images reconstructed with the two sparse regularizations ($\ell_1$+isoTV/TSV) exhibit obvious differences. In Figure~\ref{fig:raw_recimages}, we show the unconvolved (i.e., raw) reconstructed images for $\ell_1$+isoTV/TSV regularizations. As shown in Figure~\ref{fig:opt_recimages} and \ref{fig:raw_recimages}, $\ell_1$+isoTV regularization makes emission structures broader and flatter, as has already been shown in M87 simulations with the EHT \citep{akiyama2017a,akiyama2017b}. In contrast, images regularized with TSV have smoother edges with narrower photon rings (as expected from the theoretical analysis in \S\ref{sec:2}), which are closer to the ground truth images. In particular, the raw reconstructed images in Figure~\ref{fig:raw_recimages} clearly show that smooth edges in the ground truth images, which are attributed to a smooth transition in the emissivity and opacity of the plasma in the accretion flow, are much better reconstructed with $\ell_1$+TSV regularization. As a consequence of this, the TSV term comes reproduces a much clearer shadow feature in the reconstructed images. For the Free-fall model, the size of the black hole shadow is larger in the $\ell_1$+TSV image that the isoTV term and gets closer to the ground truth than the isoTV term. For sub-Keplerian and Keplerian models, the black hole shadow is visible in the $\ell_1$+TSV images but is mostly obscured (except for the darker funnel region) in the $\ell_1$+isoTV methods. 

The appearance of the reconstructed images indicates that $\ell_1$+TSV regularization is justified based on a more physically reasonable assumption and is  therefore more suitable to image the objects seen in many astronomical observations. In the following subsections, we evaluate the images more quantitatively with the image fidelity metrics described in \S\ref{sec:4}.

\subsection{NRMSE Analysis and Optimal Beam Sizes}\label{sec:5.3}
\begin{figure*}
\centering
\includegraphics[width=1.0\textwidth]{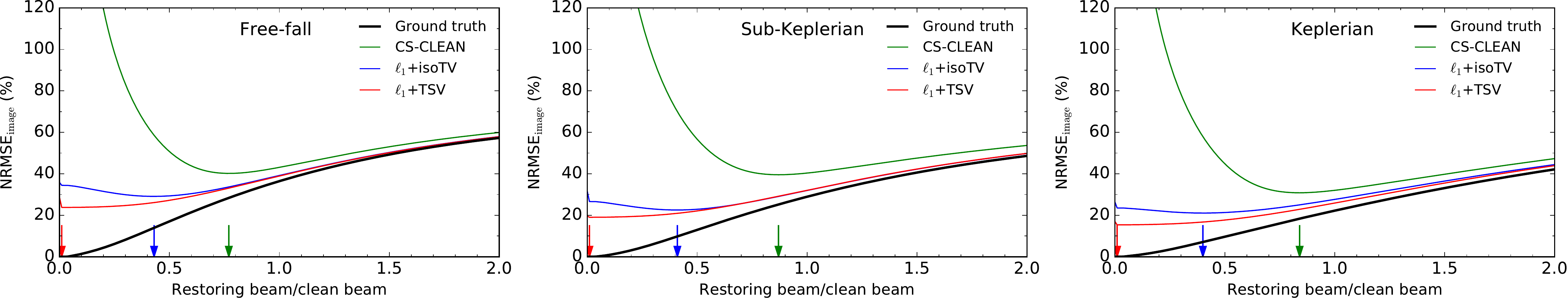}\\
(a) Image-domain NRMSE\vspace{1em}\\
\includegraphics[width=1.0\textwidth]{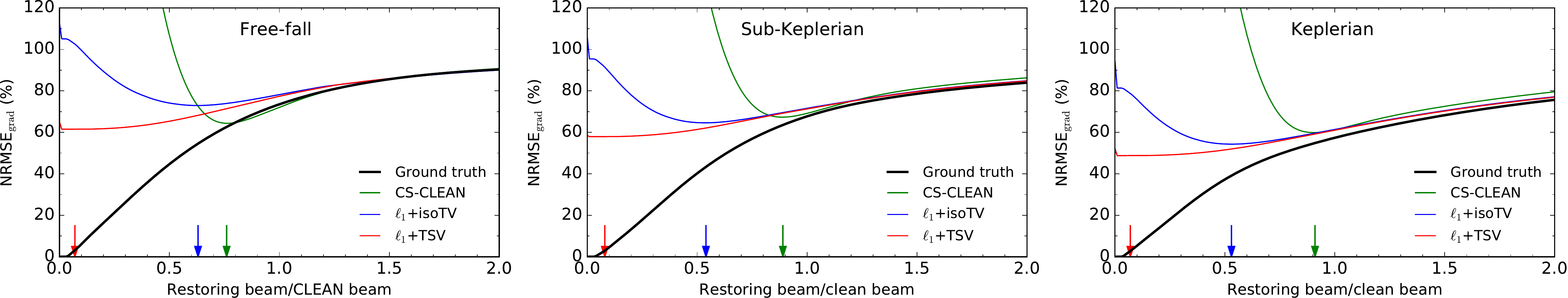}
(b) Gradient-domain NRMSE\\
\caption{The metrics of the image fidelity as measured by using NRMSE (see \S\ref{sec:4}) between the ground-truth image and ground-truth/reconstructed images convolved with the elliptical Gaussian beam.  The panel (a) shows the image domain NRMSE, while the panel (b) shows the gradient domain NRMSE. The horizontal axis shows the size of the convolving beam relative to the CLEAN beam. The green, blue and red arrows represent the optimal resolution minimizing the NRMSE value for CS-CLEAN, $\ell_1$+isoTV and $\ell_1$+TSV regularizations, respectively.}
\label{fig:nrmses}
\end{figure*}

In Figure~\ref{fig:nrmses}, we evaluate the NRMSE metric on the image domain and its gradient domain over various spatial scales, as in previous work \citep[][]{chael2016, akiyama2017a, akiyama2017b}. The black curves represent the ideal NRMSE curves between the original (unconvolved) ground truth image and the ground truth image after convolution with a Gaussian beam scaled to each resolution on the horizontal axis. These curves represent the highest fidelity available at a given resolution, as would be provided by an algorithm that reconstructs the image perfectly but at a finite (lower) resolution. The colored curves show the NRMSE between the unconvolved ground truth image and the Gaussian-convolved reconstructed images. For the NRMSE curve of the image gradient (the bottom panels), we apply the Gaussian convolution first, then take the gradient with the Sobel filter and calculate the NRMSE metric. Note that when calculating the NRMSE metrics, positional offsets between two images are corrected to minimize the NRMSE.

The behaviors of CS-CLEAN images, as seen in the image-domain NRMSE curves, is broadly consistent with previous work \citep{chael2016,akiyama2017a,akiyama2017b}.  On scales larger than the CLEAN beam, the CS-CLEAN NRMSE curves are close to the ideal curve. On scales \emph{smaller} than the CLEAN beam, the CS-CLEAN NRMSE curves rapidly deviate from the ground truth image. In the superresolution regime, CS-CLEAN reconstructions are too sparse and have too many bright, compact artifacts. The angular scale that provides the minimum NRMSE, which can be interpreted as the optimal beam size of the technique \citep{chael2016}, is $\sim 75-90$~\% of the CLEAN beam for CS-CLEAN, broadly consistent with previous work on simulations for both Sgr A* and M87 \citep[][]{chael2016,akiyama2017a,akiyama2017b}.

Sparse modeling images have much shallower curves in the image domain, consistent with our previous work on M87 simulations \citep[][]{akiyama2017a,akiyama2017b}. Both $\ell_1$+isoTV/TSV images are closer to the ideal curves than CS-CLEAN, even on scales larger than the CLEAN beam, and do not show a rapid increase in the superresolution regime like CS-CLEAN. In the superresolution regime, $\ell_1$+TSV images show smaller deviations due to smoother edges and narrower emission widths than $\ell_1$+isoTV images. Remarkably, the optimal beam size of $\ell_1$+TSV techniques is consistently $\sim 0$~\% of the CLEAN beam for all three the ground truth images, showing that a convolving beam is no longer necessary for $\ell_1$+TSV regularization. This optimal scale is much smaller than the 40-60~\% of CLEAN beam required for optimal $\ell_1$+isoTV images. Image-domain NRMSEs for unconvolved $\ell_1$+isoTV/TSV images are smaller than the reconstructed images convolved with the CLEAN beam, showing that the traditional beam-restoring process does not give better images in terms of the image-domain NRMSEs for sparse modeling techniques.

The gradient-domain NRMSE, newly introduced in this work, provides a complementary view on the performance of reconstructed images. Although the behavior and optimal beam size of CS-CLEAN images are similar to the image-domain NRMSE, differences between $\ell_1$+isoTV and $\ell_1$+TSV images are clearer in the gradient-domain NRMSE. Brightness distributions that are too flat or that have edges that are too sharp in $\ell_1$+isoTV images cause a rapid increase in the gradient-domain NRMSE curves in the superresolution regime. As a result, the NRMSEs for the non-convolved $\ell_1$+isoTV images are larger than the CLEAN-beam-convolved ones, indicating that the traditional beam-restoring process gives better images in the gradient domain. The optimal beam size of $\ell_1$+isoTV images in the gradient domain is $\sim50-70$~\% of the CLEAN beam, worse than in the image domain. Remarkably, the $\ell_1$+TSV images have similar NRMSE curves in the gradient domain as in the image domain, with an optimal beam size of $\lesssim 10$~\% of the CLEAN beam for all three models. All unconvolved $\ell_1$+TSV images have smaller NRMSE values than the CLEAN-beam convolved images, again showing that the traditional beam-restoring process is not required for $\ell_1$+TSV regularization.


\subsection{Circular Features Extracted with the CHT}\label{sec:5.4}
\begin{figure*}
	\centering
	\includegraphics[width=1.0\textwidth]{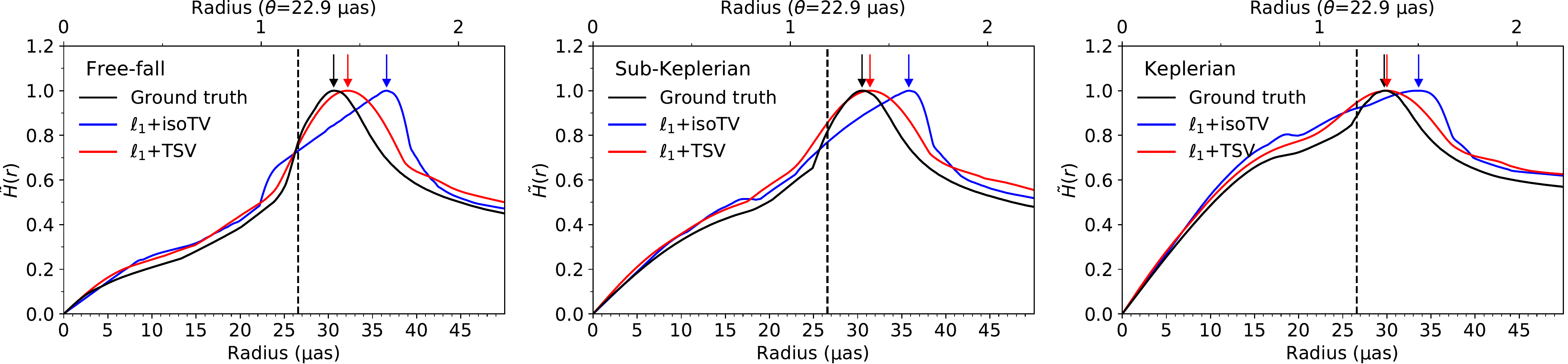} \\
	\caption{The radial profiles of the maximum CHT accumulator for the ground truth image (black line; same as in Figure~\ref{fig:rad_profile}) and the reconstructed images with $\ell_1$+isoTV (blue) and $\ell_1$+TSV (red) regularization for all three models. Each profile is normalized by its maximum value. The arrows show representative peaks of each profile, for which corresponding circles are shown in Figure~\ref{fig:raw_recimages} (see \S\ref{sec:5.4} for details).}
	\label{fig:rec_rad_profiles}
\end{figure*}
We show radial profiles of the maximum CHT accumulator for the ground truth and $\ell_1$+isoTV/TSV images in Figure~\ref{fig:rec_rad_profiles}. Here, the CHT is applied to unconvolved images in the same way as the ground truth images (see \S\ref{sec:3.2} and Figure~\ref{fig:rad_profile}), since it is practically difficult to determine the optimal resolutions in real observations where the ground truth images are unknown.


\begin{table*}
\caption{Metrics derived from the CHT}
\centering
\begin{tabular}{l l c c c c c c} \hline \hline
Metric & Unit & \multicolumn{2}{c}{Free-fall model} & \multicolumn{2}{c}{Sub-Keplerian model} & \multicolumn{2}{c}{Keplerian model} \\
& & $\ell_1$+isoTV & $\ell_1$+TSV & $\ell_1$+isoTV & $\ell_1$+TSV & $\ell_1$+isoTV & $\ell_1$+TSV \\ \hline 
Positional offset $\Delta P$ & ($\theta$)\tablenotemark{a} & 0.03 & 0.00 & 0.00 & 0.03 & 0.03 & 0.03 \\
Radius $\Delta R$ & ($R_{\rm gt}$)\tablenotemark{b} & 0.20 & 0.05 & 0.17 & 0.03 & 0.13 & 0.01 \\
& ($\theta$)\tablenotemark{a} & 0.27 & 0.07 & 0.23 & 0.04 & 0.17 & 0.01 \\ \hline 
\end{tabular}
\label{tb:metric}
\tablenotetext{a}{The diffraction limit of observations $\theta=22.9$ $\rm \mu$as.}
\tablenotetext{b}{The ground-truth peak radius.}
\end{table*}

The radial profiles of the reconstructed images (Figure~\ref{fig:rec_rad_profiles}) have a single peak around $\sim 30$~$\rm \mu$as, which is $\sim$~10\% larger than the radius of the black hole shadow. This is similar to the radial profile of the ground-truth image, which has a peak near the same location. The peak from the reconstructed images is  smoother than from the ground-truth image due to the finite angular resolution of the data. Independent of the model, $\ell_1$+isoTV images have a larger peak radius than $\ell_1$+TSV due to their flattened brightness distribution and larger emission size. As a result, peak radius can be measured more accurately with $\ell_1$+TSV regularization.  Especially for the sub-Keplerian and Keplerian models, the peak radius coincides with the ground truth. Table \ref{tb:metric} shows errors in the peak radius $\Delta R$ (see \S\ref{sec:4.2.2}) normalized to the ground-truth peak radius $R_{\rm gt}$ and also to the diffraction limit $\theta =22.9$~$\rm \mu$as. The errors are $\sim 15-20$~\% for $\ell_1$+isoTV images and $\lesssim 15$~\% for $\ell_1$+isoTV images compared to the ground-truth radius, corresponding to $\sim 35-55$~\% and $\lesssim 35$~\% of the diffraction limit, respectively.

In Figure \ref{fig:raw_recimages}, we show circles corresponding to the peaks in the radial profiles, their curvature centers, and the center of the mass of the brightness distributions. In all three models, the detected circle from the ground-truth image well represents the edge of the black hole shadow (blue circles), again demonstrating that the CHT is a useful choice for detecting circular features in images with a black hole shadow. The curvature centers of the circles almost coincides with the center of the mass of the image for the Free-fall model, with larger offsets perpendicular to the spin axis for the sub-Keplerian/Keplerian models, which have highly asymmetric images due to relativistic Doppler beaming.

$\ell _1$+TSV images show better performance in recovering not only the radii of the circles but also positional offsets between the curvature centers and the center of the mass of the image. As summarized in Table \ref{tb:metric}, the positional offset is consistent to within $3$~\% of the diffraction limit, superior to $\ell _1$+isoTV. This indicates that $\ell _1$+TSV images have better performance in reconstructing the asymmetry in images caused by relativistic Doppler beaming.

\section{Discussion}\label{sec:6}

\subsection{Total Squared Variation and Its Future Issues}
We have presented a new technique for interferometric imaging using sparse modeling with the $\ell_1$ norm and a new regularization function, Total Squared Variation (TSV). As shown in \S\ref{sec:5}, $\ell_1+$TSV regularization stably shows better performance than $\ell_1$+isoTV regularization and than the most widely used CS-CLEAN algorithm. In particular, the superiority of the combined $\ell_1$+TSV regularization is significant in the superresolution regime, where reconstruction of emission boundaries is highly dependent on the underlying assumptions of the regularization function. Our results indicate that the edge-smoothing regularization implicit in TSV is preferable to the edge-preserving characteristics of isoTV for astronomical images, which often have diffuse objects with gradual changes in the brightness distribution. 

Our results demonstrate that $\ell_1$+TSV regularization is an attractive choice for radio interferometry even compared with other sparse regularization functions.  The computational cost of TSV is cheaper than the $\ell_1$ norm of wavelet/curvelet-transformed images, which is the most popular approach in sparse reconstruction \citep{wiaux2010,wenger2010,li2011,mcewen2011,carrillo2012,carrillo2014,garsden2015,dabbech2015,onose2016,onose2017}. The high fidelity achieved with $\ell_1+$TSV regularization suggests that it is physically well matched with real astronomical images.

The technique presented in this paper can be applied to images with negative brightness distributions and is therefore also applicable to full-polarization imaging, as is $\ell_1$+isoTV regularization \citep{akiyama2017b}. Our previous work demonstrated that $\ell_1$+isoTV regularization is edge-preserving in imaging linear polarization as well as Stokes I.  In the case of linear polarization, its performance outperforms the CS-CLEAN not just in the superresolution regime but also on scales larger than the diffraction limit. These results suggest that the performance of polarimetric imaging may be improved with $\ell_1+$TSV regularization, too. In a forthcoming paper, we will evaluate the performance of $\ell_1+$TSV regularization in full-polarimetric imaging.

The edge-smoothing property of TSV would be useful to expand the current framework of sparse reconstruction from two-dimensional images to three-dimensional imaging, where the third dimension is, for instance, time (i.e., reconstruction of movies). Our imaging techniques can be applied to three-dimensional imaging problems by reconstructing each time frame of the image separately. However, if frame-to-frame variations of a three-dimensional image can be regularized with a reasonable function, the frames of a three-dimensional image can be solved simultaneously. A clear benefit of this approach is that since each frame of the image can be related to other frames of the image, it may lead to further noise suppression and higher fidelity. Since the dynamics and spectrum of the diffuse medium are usually expected to be gradual and without strong edges, TSV would be a reasonable choice to regularize the brightness distribution along the third dimension. Recently, \citet{johnson2017} presented movie reconstruction techniques, including a technique using MEM regularization for two-dimensional image and TSV regularization for time direction, and demonstrated that TSV can be a good regularization function in the time direction.  Such an extension to higher dimensions will enable our techniques to handle intra-day time-variable sources like Sgr A* and also high-fidelity multi-epoch imaging.

\subsection{Implications for the Near Future Observations of Sgr A* and M87 with the EHT}
With simulated observations, the black hole shadow of Sgr A* is clearly reproduced with $l_1+$TSV regularization for all tested models. Combined with the results of simulations of M87 in our previous work \citep{ikeda2016, akiyama2017a}, this work demonstrates that near-future EHT observations will have sufficient $uv$-coverage and sensitivity to image the silhouette of the black hole and surrounding magnetized plasma on $R_s$ scales in both sources.

Furthermore, we propose a feature extraction method using the CHT to detect circular features in an image, and we evaluate its performance on ground-truth and reconstructed images. Application of the CHT (see Figures \ref{fig:rad_profile} and \ref{fig:raw_recimages}) demonstrates that the CHT can extract a circle of $\sim 30~\rm \mu$as, representing the photon ring itself, regardless of the degree of relativistic Doppler-beaming for semi-analytic RIAF models \citep{broderick2006,broderick2011a,broderick2011b,broderick2016,pu2016} and also of the ground-truth and reconstructed images. 
Considering that its angular radius is $\sim 27$ $\mu$as for the non-rotating black hole case, the shadow angular size can be constrained with an accuracy of $\sim 10-20$\% using circles detected with the CHT. Although we present only the CHT of the unfiltered images, another constraint would be able to be obtained from the CHT of the shadow region of the image by using the gradient or Canny filter as demonstrated in \citet{psaltis2015}. More detailed study of the application of the CHT could lead to further improvements in the accuracy of the physical parameter estimations.

A measurement of the size of the black hole shadow can be used to estimate the mass of the SMBH. This measurement would be complementary to other methods of black-hole mass measurements, such as using the orbits of circumnuclear stars for Sgr A* where each star can be resolved \citep[e.g.~see][and reference therein]{ghez2008,chatzopoulos2015}, emission lines from circumnuclear stars \cite[e.g.][]{gebhardt2011} and gas \citep[e.g.][]{walsh2013} for M87 and nearby galaxies where stars cannot be resolved. In particular, for M87 and nearby galaxies, recent studies suggest that there is a systematic difference in the measured black hole mass by a factor of $\sim$2 between stellar- and gas-dynamical models \citep[see][]{walsh2013}. The achieved accuracy of $\sim 10-20$~\% is more than sufficient to resolve a factor-of-two discrepancy. Therefore, measurements of the circular features in the images of Sgr A* and M87 would clarify which modeling method is preferable for measuring the mass of SMBHs in nearby galaxies. This is an important input for recalibrating the $M-\sigma$ relation between the stellar velocity dispersion $\sigma$ of a galaxy bulge and the mass $M$ of the SMBH and is therefore useful to understand the coevolution of the SMBH and its host galaxy \citep{kormendy2013}.

A relevant item for future work is applying these feature extraction methods to more diverse models for Sgr A* and M87, in particular simulated with time-variable general-relativistic magnetohydrodynamics (GRMHD) \citep[e.g.][]{dexter2012,chan2015a,chan2015b,moscibrodzka2014,moscibrodzka2017},  over a wide range of physical parameters. The relation between the derived parameters of the detected circles and more fundamental physical parameters such as the black hole spin and mass would be an interesting issue for future work. The EHT is still expanding and is expected to have both higher sensitivities and more stations in the next several years than were assumed for the simulated observations in this work. Feature extraction with the CHT will be useful to evaluate the impact of potential future improvements in the EHT array.

\section{Conclusion}
We have presented a new imaging technique for radio/optical interferometry to obtain high-fidelity superresolution images using two sparse regularizers: the $\ell_1$ norm and a new function named Total Squared Variation (TSV). The edge-smoothing feature of TSV regularization is physically reasonable for the astronomical images which are often diffuse and without strong edges. As an example, we apply the proposed technique to simulated observations of Sgr A* with the Event Horizon Telescope at 1.3~mm (230~GHz). To evaluate the reconstructed images, we develop new metrics using the image gradient and a feature-extraction method using the circle Hough transform (CHT), both of which are inspired by recent theoretical work on Sgr A* in \cite{psaltis2015}. These methods provide new ways to inspect and analyze the reconstructed images from a more physical point of view. 

We summarize our main conclusions as follows.
\begin{enumerate}
\item With geometric models consisting of two point sources, we demonstrate that both $\ell_1$+isoTV and $\ell_1$+TSV can separately identify two components with an interval of $\sim$30\,\% of the CLEAN beam size, which is therefore the effective angular resolution of the techniques. These results are consistent with previous work that new imaging techniques utilizing convex regularization functions generally provide a capability of superresolution of $\sim$0.25 $\lambda /D$. The actual resolution obtained may be worse when the quality of the data is poor.
\item Our new $\ell_1$+TSV regularization successfully reconstructs the black hole shadow for all three models. It outperforms both $\ell_1$+isoTV regularization and the Cotton-Schwab CLEAN algorithm. Remarkably, the optimal beam size achieved with $\ell_1$+TSV regularization is $\lesssim 10$~\% of the diffraction limit, and unconvolved reconstructed images have smaller errors than images that have undergone post-imaging beam convolution.  This indicates that the traditional method of the Gaussian convolution with a restoring beam in interferometric imaging would be no longer required for the $\ell_1$+TSV regularization.  
\item The CHT is an attractive method to extract circular features from the images with the black hole shadow. We find that the CHT can constrain the radius of the black hole shadow to an accuracy of $\sim 10-20$~\% for reconstructed images with $\ell_1$+TSV regularization. This would provide complementary verification of current mass measurements of the SMBHs in Sgr A* and M87.  
\end{enumerate}
These results demonstrate that $\ell_1$+TSV regularization is an attractive choice for superresolution interferometric imaging, which can provide an unprecedented view of the event-horizon-scale structure of the super-massive black hole in Sgr A* and M87 with the EHT. \vspace{1eM}

\acknowledgements
This work is financially supported by the MEXT/JSPS KAKENHI (Grant Numbers 24540242, 25120007 and 25120008) and also a grant from the National Science Foundation (AST-1614868). K.A. is financially supported by the program of the Jansky Fellowship of the National Radio Astronomy Observatory and the above grant from the NSF. Event Horizon Telescope work at MIT Haystack Observatory is supported by the grants from the NSF (AST-1440254, AST-1614868) and through an award from the Gordon and Betty Moore Foundation (GMBF-3561). The National Radio Astronomy Observatory is a facility of the National Science Foundation operated under cooperative agreement by Associated Universities, Inc. The Black Hole Initiative at Harvard University is financially supported by a grant from the John Templeton Foundation. This research made use of Astropy, a community-developed core Python package for Astronomy \citep{astropy2013}.

\software{MAPS, CASA, astropy}

\bibliographystyle{yahapj}

\end{document}